\documentclass[
]{ceurart}

\usepackage{booktabs}
\usepackage{multirow}
\usepackage{amsmath}
\usepackage{xurl}

\begin{document}
	
	\copyrightyear{2026}
	\copyrightclause{Copyright for this paper by its authors.
		Use permitted under Creative Commons License Attribution 4.0 International (CC BY 4.0).}
	\conference{GeCoIn 2026: Generative Code Intelligence Workshop, co-located with the 35th International Joint Conference on Artificial Intelligence (IJCAI-ECAI 2026),  August 15--17, 2026 --- Bremen, Germany}
	
	\title{Loc2Repair: A Framework for Evaluating the Impact of File-Level Issue Localization in Repo-Level LLM Repair}
	
	\author[1]{Mohammad Nour Al Awad}[
	email=MohammadNourAlAwad@itmo.ru
	]
	
	\author[1]{Sergey Ivanov}[
	email=svivanov@itmo.ru
	]
	
	\address[1]{ITMO University}
	
	\begin{abstract}
		Repository-grounded automated repair is often reported as a single end-to-end capability, which hides distinct failure modes such as poor file targeting, incorrect patch synthesis, and failed iterative debugging. We present \emph{Loc2Repair}, a modular evaluation framework for controlled analysis of repository-grounded repair pipelines, and use it to isolate file-level issue localization as an upstream variable. Loc2Repair decouples localization and repair under a shared runtime, artifact schema, and evaluation harness, allowing researchers to combine different localization models and repair backbones under matched conditions. Using three repair backbones on SWE-bench Verified, we compare baseline repair without explicit localization, repair guided by predicted localization from two localizers, and repair guided by gold modified-file sets. Explicit localization consistently improves resolved rate across all backbones: pooled performance increases from \(44.7\%\) for baseline repair to \(48.9\%\) and \(49.1\%\) with predicted localization, and to \(52.4\%\) with gold localization. Localization also reduces mean elapsed time overall: in pooled paired analysis, mean elapsed time decreases by \(100.94\)s and \(52.25\)s for the two predicted-localization settings, and by \(154.45\)s with gold guidance, although token effects remain heterogeneous across models. Overall, Loc2Repair shows file-level localization is a consistent repair lever, improving effectiveness and mean latency in pooled analysis, while gold-guided failures expose headroom beyond localization.
	\end{abstract}
	
	
	\maketitle
	
	\section{Introduction}
	
	Repository-grounded software-engineering agents now operate over full repositories, inspect files, execute commands, run tests, and propose patches for realistic issues \cite{SWEbench2024,SWEagent2024,Agentless2025,AutoCodeRover2024,OpenHands2025,SGAgent2026}. This shift has made end-to-end issue resolution a central evaluation target for large language model (LLM) systems. However, end-to-end reporting alone is often too coarse for scientific diagnosis. When a repair attempt fails, the cause may lie in at least three distinct places: the agent may search the wrong region of the repository, synthesize an incorrect patch after reaching the right files, or fail to recover through iterative validation and debugging.
	
	For repository-grounded repair, file targeting is a particularly important upstream decision. Before modifying code, an agent must decide what to inspect, which files are likely relevant, and how to allocate limited attention over a repository. Errors at this stage propagate downstream by distorting context selection, tool use, and repair strategy. As a result, weak localization can suppress downstream repair performance even when the repair backbone itself is capable of producing a correct patch once given the right starting context.
	
	This paper introduces \emph{Loc2Repair}, a modular evaluation framework for controlled analysis of repository-grounded repair pipelines. Loc2Repair decouples localization and repair under a shared runtime, artifact schema, and evaluation harness, allowing researchers to combine different localization models and repair backbones, compare them under matched conditions, and quantify both effectiveness and cost. In this paper, we use the framework to study one central question: \emph{how much does file-level issue localization change downstream repair when the repair backbone is otherwise held fixed?}
	
	Concretely, for each repair backbone, we compare four conditions: baseline repair without explicit localization, repair guided by predicted localization from Qwen4B, repair guided by predicted localization from Gemma4E4B, and repair guided by gold modified-file sets derived from benchmark references. This design isolates localization as an experimental variable while simultaneously providing a reusable evaluation framework for future plug-and-play comparisons of new localizers and repair models.
	
	We study four research questions:
	\begin{itemize}
		\item \textbf{RQ1.} Does explicit file-level localization improve downstream repair effectiveness?
		\item \textbf{RQ2.} How consistent are localization-induced gains across repair backbones?
		\item \textbf{RQ3.} What efficiency trade-offs does localization introduce?
		\item \textbf{RQ4.} What bottlenecks remain beyond file-level localization?
	\end{itemize}
	
	Across three repair backbones, explicit localization consistently improves resolved rate, while gold localization improves further and reveals headroom. Pooled analysis shows reliable aggregate gains and lower mean elapsed time, although token effects remain model-dependent. This combination of effectiveness and efficiency mirrors a broader lesson from LLM code-assistance research: operational choices around when model assistance is invoked or presented can affect both usefulness and wasted inference~\cite{AlAwadTiming2025}. 
		
	The paper makes four contributions:	
	\begin{itemize}
		\item We present \textbf{Loc2Repair}, a modular framework that decouples localization and repair under a shared runtime and evaluation harness, enabling controlled comparisons across localizers and repair backbones.
		\item We provide cross-backbone evidence on SWE-bench Verified that explicit file-level localization consistently improves end-to-end repair effectiveness.
		\item We show that localization can improve effectiveness and latency: pooled analysis yields resolved-rate gains and lower mean elapsed time, while token effects remain model-dependent.
		\item We derive evaluation implications for repository-grounded code agents, showing how stage-aware analysis exposes headroom that end-to-end reporting alone would hide.
	\end{itemize}
	\section{Related Work}

	\paragraph{Repository-grounded repair.}
	Recent benchmarks, datasets, and systems established realistic evaluation for issue resolution in full software repositories \cite{SWEbench2024,SWEagent2024,Agentless2025,AutoCodeRover2024,RepairAgent2024,OpenHands2025,ThinkRepair2024,FixAgent2024,RepoDebug2025,SWEPolyBench2025,SGAgent2026}. This line of work shifted repair toward realistic agentic settings with repository navigation, tool use, execution, validation, memory, and repair-data generation in the loop \cite{ExpeRepair2025,SWESynth2025}. Cost-aware invocation has also been studied in developer-facing LLM coding tools, where behavioral telemetry can suppress low-value model calls before generation~\cite{AlAwadPrefiltering2025}. At the same time, it made attribution harder: end-to-end success or failure does not by itself reveal whether a system failed because it searched the wrong files, generated the wrong patch, or failed during iterative repair.
	
	\paragraph{Bug and issue localization.}
	Bug and issue localization have a long history in software engineering. Classical approaches emphasized information retrieval, ranking, and hybrid signals that combine textual and structural evidence \cite{SahaLKP2013ASE_BLUiR,YeBL2014FSE_LTR,WangLo2016AmalgamPlus,AkbarK2020MSR_Ablation,Niu2025CSUR_DLmeetsIR}. More recent work has extended localization to the LLM era, including repository-level issue localization, graph-guided localization, and agentic localization strategies \cite{LocAgent2025,CoSIL2025,OrcaLoca2025,GottaCatchEmAll2025,SweRank2026}. These works strengthen localization itself, but they do not directly answer a separate intervention question that matters for repository-grounded repair: how much downstream repair quality changes when file-level localization quality changes while the repair backbone is otherwise held fixed.
	
	\paragraph{Localization and repair together.}
	A growing body of work combines localization and repair within integrated LLM-based pipelines \cite{Hossain2024DeepDive,DEVLoRe2025,RGFL2026}. This is an important systems direction, since realistic repair agents must ultimately coordinate search, reasoning, editing, and validation. However, integrated architectures make stage attribution difficult: when end-to-end performance improves, it is often unclear how much of the gain comes from stronger localization versus stronger downstream repair behavior.
	
	Our contribution is complementary to these directions. We perform a controlled intervention study that varies file-level localization while keeping the downstream repair setting fixed. In that sense, the paper is not only about repair performance, but also about \emph{evaluation methodology}: it asks how stage-aware interventions can help explain where repository-grounded repair systems gain or lose performance.
	
	\section{Loc2Repair Framework and Study Design}
	
	\subsection{Design Principle: Vary Localization, Hold Repair Fixed}
	
	The central goal of this study is attribution. End-to-end repair quality is affected by many interacting components, so a clean analysis requires manipulating one stage while leaving the downstream repair regime unchanged. We therefore treat file-level localization as an explicit intervention variable.
	
	For each repair backbone, we compare four arms:
	\begin{itemize}
		\item \textbf{Baseline}: repair without an explicit localization stage.
		\item \textbf{Pred-Qwen4B}: repair with file hints predicted by Qwen4B localization.
		\item \textbf{Pred-Gemma4E4B}: repair with file hints predicted by Gemma4E4B localization.
		\item \textbf{Gold}: repair with oracle-style file hints derived from benchmark reference modified-file sets.
	\end{itemize}
	
	This structure yields two complementary comparisons. Predicted-localized repair quantifies the practical value of using smaller localization models as an upstream guidance stage for larger repair backbones. We use Qwen4B and Gemma4E4B as two representative lightweight localizers with different quality--cost profiles, testing whether compact upstream guidance can improve larger repair backbones without changing the repair backbone itself. Gold-localized repair provides an oracle-style intervention, estimating remaining headroom under substantially stronger file targeting.
	
	\subsection{Framework Architecture and Runtime Pipeline}
	
	Loc2Repair is designed as a reusable framework rather than a one-off script for a single model pairing. It exposes localization and repair as separate, swappable stages under a shared runtime, artifact schema, and benchmark-normalized evaluation pipeline. This makes it possible to plug in new localizers or repair backbones, rerun the same protocol, and compare combinations under matched repository snapshots, prompts, telemetry fields, and output normalization.
	
	At runtime, Loc2Repair executes each task using a mini-SWE-agent-based repository interaction loop \cite{MiniSWEAgent2025}. Each task begins from a repository snapshot and issue description exported from SWE-bench Verified. A run initializes an isolated workspace, materializes the repository state for that instance, and then executes the configured stages. For repair, mini-SWE-agent is allowed to inspect files, edit the workspace, execute commands, and submit a final patch. For localization, we use the same agent scaffold with different prompts and read-only constraints: it may inspect the repository, but it must return only a structured list of predicted files.
	
	The framework stores structured artifacts for each stage, including repair outcomes, cost telemetry, localization outputs, and patch metadata. This shared schema supports controlled cross-combination analysis: the same harness can evaluate localization--repair pairings not only by resolved outcomes, but also by operational metrics such as runtime and token usage across matched instances.
	
	\subsection{Localization Stage and Repair Injection}
	At the localization stage, the agent receives the issue text, repository workspace, and instance metadata. It is prompt-constrained to inspect files and directory structure, not to edit the repository, and to return only a structured list of predicted files. The localization workspace is disposable: only the prediction JSON and trace are retained, so accidental edits cannot affect repair. The stage must return a single terminal JSON object of the form: \texttt{\{"predicted\_files":["repo/relative/path", ...]\}}
	
	Localization outputs are validated and normalized before being used as repair guidance. We use \emph{soft localization}: predicted or gold file sets are injected into the repair prompt as high-confidence starting targets, but are not enforced as hard edit constraints. Thus, the repair agent can still inspect other files, reason over the wider repository, and decide how to proceed.
	
	Within each repair-backbone comparison, the arms differ only in their initial file guidance: no explicit hints for baseline, predicted file hints for the predicted-localized arms, and reference modified-file hints for the gold arm. The repair backbone, execution environment, and evaluation harness remain fixed.
	
	Table~\ref{tab:loc_pooled} reports standalone localization quality and cost. Each localizer was executed three times on the same 500 instances, and we report all six artifacts directly rather than pooling them. Gemma4E4B is assigned a larger timeout because it is a slower localizer. Later runtime and token measurements are end-to-end: predicted-localized arms include localization and repair, baseline includes repair only, and gold uses reference file hints without localizer inference.
	
	\begin{table}[t]
		\centering
		\caption{Standalone localization quality and cost for cached artifacts used downstream.}
		\label{tab:loc_pooled}
		\scriptsize
		\setlength{\tabcolsep}{2pt}
		\begin{tabular}{llccccccccc}
			\toprule
			\textbf{Localizer} & \textbf{Artifact} & \textbf{Completed} & \textbf{Limit} & \textbf{Timeout} & \textbf{F1} & \textbf{Exact} & \textbf{Hit@1} & \textbf{MRR} & \textbf{Time (s)} & \textbf{Tokens (in/out)} \\
			\midrule
			Qwen4B & A & 491/500 & 90s & 1.6\% & 57.0\% & 48.0\% & 61.0\% & 62.0\% & 18.37 & 13{,}389 / 624 \\
			Qwen4B & B & 495/500 & 90s & 0.6\% & 58.0\% & 48.0\% & 62.0\% & 63.0\% & 15.25 & 10{,}730 / 577 \\
			Qwen4B & C & 498/500 & 90s & 0.4\% & 59.0\% & 49.0\% & 62.0\% & 64.0\% & 15.25 & 11{,}036 / 565 \\
			Gemma4E4B & A & 493/500 & 180s & 1.4\% & 60.0\% & 53.0\% & 64.0\% & 65.0\% & 59.06 & 32{,}289 / 1{,}851 \\
			Gemma4E4B & B & 487/500 & 180s & 2.6\% & 61.0\% & 54.0\% & 65.0\% & 65.0\% & 69.49 & 60{,}618 / 1{,}997 \\
			Gemma4E4B & C & 478/500 & 180s & 4.4\% & 60.0\% & 53.0\% & 63.0\% & 64.0\% & 70.72 & 99{,}657 / 2{,}117 \\
			\bottomrule
		\end{tabular}
	\end{table}
	
	All runs are executed in isolated sandboxes with fixed CPU, memory, and timeout limits. These controls keep repair comparable across arms and prevent unbounded resource use. Localization artifacts were cached separately for each downstream experiment block, so cross-backbone comparisons include localization-artifact variation. This does not affect the primary within-backbone paired comparisons, but limits pooled interpretation.
	
	\section{Experimental Setup}
	
	\subsection{Dataset and Models}
	We evaluate each repair backbone on the same 500 SWE-bench Verified instances. The three repair backbones are Gemma4-26B-A4B-IT, GLM-4.7, and Qwen3.5-35B-A3B. For each backbone, we evaluate the four arms described above: baseline, Pred-Qwen4B, Pred-Gemma4E4B, and gold localization. Pooled analyses aggregate results across the three repair backbones at the backbone--instance level and are reported as aggregate paired evidence complementary to the backbone-specific comparisons. In our experiments, localizer and repair model endpoints were served with vLLM on internal GPU servers using 4$\times$ NVIDIA RTX 6000 Ada GPUs per deployed model. Reported elapsed-time measurements include model inference latency under this serving configuration.

	\subsection{Outcome Definition}
	
	Primary effectiveness is measured by the standard SWE-bench notion of \emph{resolved}: whether the generated patch passes the benchmark evaluation harness for the given instance. For each repair backbone, all arms are evaluated on the same 500 SWE-bench Verified instances. Runs that do not produce a valid final patch within the 30-minute repair budget are counted as unresolved. This keeps the denominator fixed and makes comparisons across arms consistent.
	
	The secondary outcomes are average runtime and average token usage. Runtime captures operational latency, while tokens provide a model-centric measure of computational cost. Together, these outcomes show whether effectiveness gains come with favorable, neutral, or adverse efficiency trade-offs.
	
	\subsection{Pairing and Statistical Analysis}
	
	All comparisons are paired by benchmark instance within each backbone. For resolved outcomes, we use exact McNemar tests on discordant outcomes and report paired 95\% confidence intervals for resolved-rate deltas. Since the same 500 instances are reused across backbones, pooled results are treated as aggregate evidence.
	
	For efficiency, we compute per-instance deltas between each localized arm and the corresponding baseline over instances with numeric telemetry in both arms. We use two-sided sign tests over non-tied paired differences. Throughout the paper, negative efficiency deltas mean that the localized arm is better, i.e., faster or cheaper. We report raw \(p\)-values and use \(p<0.05\) as the significance threshold, without multiple-testing correction.
	
	\section{Results}
	
	\subsection{RQ1: Localization improves repair effectiveness consistently}
	Table~\ref{tab:main_results} reports the main effectiveness and efficiency outcomes for all three backbones. Resolved outcomes use the same 500 SWE-bench Verified instances per backbone. Runtime and token values are descriptive arm-level means, with token counts rounded to thousands; paired efficiency tests are reported in Table~\ref{tab:paired_summary}.
	
	\begin{table}[t]
		\centering
		\caption{Main repair results by backbone and intervention arm.}
		\label{tab:main_results}
		\scriptsize
		\setlength{\tabcolsep}{4pt}
		\begin{tabular}{llcccc}
			\toprule
			\textbf{Backbone} & \textbf{Metric} & \textbf{Baseline} & \textbf{Pred-Qwen4B} & \textbf{Pred-Gemma4E4B} & \textbf{Gold} \\
			\midrule
			\multirow{4}{*}{Gemma4}
			& Resolved & 216/500 (43.2\%) & 239/500 (47.8\%) & 226/500 (45.2\%) & 249/500 (49.8\%) \\
			& Runtime (s) & 662.50 & 592.10 & 592.87 & 521.49 \\
			& Loc. tok. (in/out, K) & 0.0 / 0.0 & 7.1 / 0.5 & 29.1 / 1.8 & 0.0 / 0.0 \\
			& Repair tok. (in/out, K) & 300.5 / 8.4 & 356.0 / 10.1 & 413.8 / 9.9 & 346.4 / 10.4 \\
			\midrule
			\multirow{4}{*}{GLM-4.7}
			& Resolved & 187/500 (37.4\%) & 206/500 (41.2\%) & 218/500 (43.6\%) & 233/500 (46.6\%) \\
			& Runtime (s) & 918.88 & 963.70 & 1045.30 & 947.71 \\
			& Loc. tok. (in/out, K) & 0.0 / 0.0 & 7.5 / 0.5 & 26.2 / 1.7 & 0.0 / 0.0 \\
			& Repair tok. (in/out, K) & 1{,}332.3 / 13.5 & 1{,}291.8 / 13.0 & 1{,}305.0 / 13.4 & 1{,}365.8 / 13.9 \\
			\midrule
			\multirow{4}{*}{Qwen3.5}
			& Resolved & 267/500 (53.4\%) & 289/500 (57.8\%) & 293/500 (58.6\%) & 304/500 (60.8\%) \\
			& Runtime (s) & 935.88 & 816.13 & 874.49 & 799.66 \\
			& Loc. tok. (in/out, K) & 0.0 / 0.0 & 7.3 / 0.5 & 25.0 / 1.7 & 0.0 / 0.0 \\
			& Repair tok. (in/out, K) & 1{,}224.0 / 9.6 & 1{,}228.9 / 10.4 & 1{,}185.9 / 10.3 & 1{,}212.5 / 10.4 \\
			\bottomrule
		\end{tabular}
	\end{table}
	
	Relative to baseline, predicted localization increases resolved rate by \(+4.6\) and \(+2.0\) percentage points for Gemma4, \(+3.8\) and \(+6.2\) points for GLM-4.7, and \(+4.4\) and \(+5.2\) points for Qwen3.5. Gold localization yields the largest gains in every backbone: \(+6.6\), \(+9.2\), and \(+7.4\) points, respectively.
	
	Pooled across the three repair backbones, baseline repair resolves \(44.7\%\) of cases. Pred-Qwen4B increases this to \(48.9\%\), Pred-Gemma4E4B to \(49.1\%\), and gold localization to \(52.4\%\).	Thus, explicit localization produces a non-trivial lift before statistical testing, and oracle-style localization lifts performance further. At the same time, Table~\ref{tab:loc_pooled} shows that stronger standalone localization does not translate monotonically into larger downstream repair gains: Gemma4E4B is modestly stronger as a localizer, but its downstream advantage over Qwen4B is small in aggregate and not uniform across backbones.
	
	\subsection{RQ2: The gains are positive across backbones, and gold localization reveals clear headroom}
	
	Table~\ref{tab:paired_summary} summarizes paired comparisons against baseline. Resolved-rate deltas are in percentage points; negative efficiency deltas mean the localized arm is faster or cheaper. Predicted localization is positive in all cases and significant in four of six backbone-level comparisons; gold is significant for all backbones.
	
	\begin{table*}[t]
		\centering
		\caption{Paired comparisons against baseline.}
		\label{tab:paired_summary}
		\scriptsize
		\setlength{\tabcolsep}{4pt}
		\resizebox{\textwidth}{!}{%
		\begin{tabular}{llccccc}
		\toprule
		\textbf{Backbone} & \textbf{Comparison} & \textbf{\(\Delta\) Resolved, \(p\)} & \textbf{95\% CI} & \textbf{\(\Delta\) Time, \(p\)} & \textbf{\(\Delta\) Input Tok. (K), \(p\)} & \textbf{\(\Delta\) Output Tok. (K), \(p\)} \\
			\midrule
			Gemma4 & Pred-Qwen4B & +4.6 (0.033) & [+0.6, +8.6] & -171.4, 5.4e-08 & +61.2, 0.016 & +2.1, 3.6e-05 \\
			Gemma4 & Pred-Gemma4E4B & +2.0 (0.387) & [-2.1, +6.1] & -158.5, 3.8e-06 & +128.1, 0.0017 & +3.3, 3.7e-16 \\
			Gemma4 & Gold & +6.6 (0.0015) & [+2.7, +10.5] & -250.1, 1.4e-20 & +50.4, 0.858 & +2.0, 0.00091 \\
			\midrule
			GLM-4.7 & Pred-Qwen4B & +3.8 (0.110) & [-0.6, +8.2] & -0.9, 0.396 & -195.0, 0.116 & -0.5, 0.262 \\
			GLM-4.7 & Pred-Gemma4E4B & +6.2 (0.0057) & [+2.0, +10.4] & +71.6, 0.00020 & -161.6, 0.0061 & +0.7, 0.0035 \\
			GLM-4.7 & Gold & +9.2 (8.7e-05) & [+4.7, +13.7] & -53.8, 0.044 & -216.8, 0.0040 & -0.6, 0.072 \\
			\midrule
			Qwen3.5 & Pred-Qwen4B & +4.4 (0.043) & [+0.3, +8.5] & -130.5, 0.00020 & +106.9, 0.067 & +1.8, 1.1e-08 \\
			Qwen3.5 & Pred-Gemma4E4B & +5.2 (0.016) & [+1.2, +9.2] & -69.9, 0.396 & +86.1, 0.044 & +2.7, 5.8e-16 \\
			Qwen3.5 & Gold & +7.4 (0.00045) & [+3.4, +11.4] & -159.4, 3.9e-10 & +48.9, 0.195 & +1.1, 0.0083 \\
			\bottomrule
		\end{tabular}
		}
	\end{table*}
	
	Pooled paired tests show the same direction, where baseline \(\rightarrow\) Pred-Qwen4B yields \(+4.3\) points (95\% CI \([+1.9,+6.7]\), \(p=0.0006365\)); baseline \(\rightarrow\) Pred-Gemma4E4B yields \(+4.5\) points (95\% CI \([+2.1,+6.8]\), \(p=0.0002982\)); baseline \(\rightarrow\) gold yields \(+7.7\) points (95\% CI \([+5.3,+10.1]\), \(p=3.874\times 10^{-10}\)). The discordant-pair counts point in the same direction: for example, in the pooled gold comparison, 230 instances flip from unresolved under baseline to resolved under gold localization, versus only 114 moving in the opposite direction.
	
	Together, these findings support an intervention-based interpretation: changing file guidance under a fixed repair backbone systematically changes outcomes. Current predicted localization also leaves substantial value relative to gold guidance.
	
	\subsection{RQ3: Efficiency effects are real, but they are not uniform}
\begin{figure}[t]
		\centering
		\includegraphics[width=\linewidth]{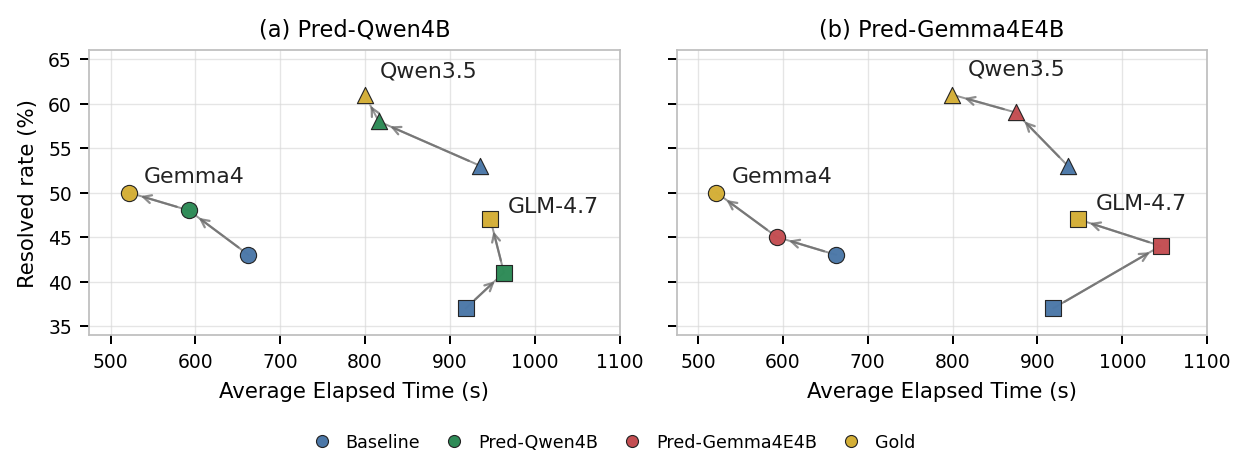}
		\caption{Resolved rate versus average elapsed time by repair backbone. Upper-left is better; arrows indicate the effectiveness--latency shift under localization.}
		\label{fig:resolved_vs_time}
	\end{figure}

	Localization is not only an effectiveness intervention; it often improves runtime as well. Figure~\ref{fig:resolved_vs_time} shows the arm-level effectiveness--latency trade-off, where upper-left is better. Gemma4 and Qwen3.5 are faster under all settings, while GLM-4.7 is the main counterexample: predicted localization improves effectiveness but can increase elapsed time. One plausible explanation is that localization helps GLM-4.7 complete harder instances that baseline repair either fails or times out on. Token usage is also mixed, so efficiency depends on the backbone--localizer pairing rather than following a single global rule.

	The pooled paired means summarize the overall balance: Pred-Qwen4B yields \(-100.94\)s and \(-7{,}377\) tokens; Pred-Gemma4E4B yields \(-52.25\)s and \(+20{,}187\) tokens; gold yields \(-154.45\)s and \(-37{,}835\) tokens. Thus, pooled analysis shows that localization raises resolved rate and reduces elapsed time on average, although token effects remain model-dependent.
	
	\subsection{RQ4: Better localization improves repair, but it does not saturate it}
	
	The clearest evidence that localization is only part of the problem comes from the gold arm. Even when repair is initialized with the benchmark's reference modified-file sets, unresolved counts remain substantial: 251 for Gemma4, 267 for GLM-4.7, and 196 for Qwen3.5.
	
	This matters because gold localization sharply reduces file-targeting error as a front-end bottleneck. If file targeting were the only major residual failure mode, gold guidance should saturate much more. Instead, the remaining unresolved volume is too large to ignore. The implication is direct: repository-grounded repair still depends heavily on downstream factors such as semantic patch synthesis, multi-step debugging, validation behavior, and broader repair-time decision policy. This suggests that future repair systems may benefit from decomposing repair into more than two stages, rather than treating localization and patch generation as the only separable components.

	\section{Discussion}
	
	\subsection{Localization is a first-class performance lever}
	The primary result is not merely that localization helps, but that it helps consistently. Predicted-localization gains are positive across all backbones, while gold localization is significant for all three. 
	
	At the same time, the gold condition argues against an overly narrow narrative in which localization is \emph{the} bottleneck. Gold guidance improves all backbones further than predicted localization, but still leaves many instances unresolved. This residual gap suggests that repository-grounded repair is a staged problem involving file targeting, semantic patch synthesis, iterative debugging, validation behavior, and broader repair-time policy.
		
	\subsection{Implications for code-agent design}
	Treating file targeting as implicit emergent behavior hides a controllable source of variance. A modular localization stage can improve end-to-end repair without changing the repair backbone itself, and gold-localization headroom suggests that investing in localizer quality remains useful.
	
	A broader implication is that repair pipelines may benefit from further decomposition beyond localization and patch generation. Tasks such as context selection, patch planning, edit synthesis, test selection, failure diagnosis, and patch refinement may be handled by specialized stages or specialized models. Such decomposition could improve coordination, keep contexts cleaner, and allow smaller models to handle narrower subtasks, potentially reducing cost while improving reliability.
	
	Localization should be optimized jointly for quality and cost. Our results show that better resolved rate does not imply uniformly lower runtime or token usage. Practical deployments should therefore evaluate localization under cost-adjusted utility rather than accuracy alone. 
	
	\subsection{Implications for evaluation methodology}
	
	The findings also have implications for benchmarking. End-to-end leaderboards remain necessary because they capture actual user-visible performance. But they are insufficient for diagnosis. Two systems with similar resolved rates may differ sharply in where they succeed or fail: one may localize well and repair poorly, while another may localize poorly but patch well when correctly guided.
	
	Stage-aware reporting can expose those differences. In our setting, paired discordant-outcome counts, standalone localization metrics, gold-localization interventions, and efficiency deltas together provide a much more informative picture than aggregate resolved rate alone. In particular, the standalone localization results show that the strongest localizer in isolation is not automatically the strongest repair partner downstream, which would be invisible in a pure end-to-end leaderboard. We therefore view oracle-style stage interventions and plug-and-play stage evaluation not as artificial shortcuts, but as useful diagnostic tools for understanding where future gains are likely to come from.
	
	\section{Threats to Validity}
	
	\textbf{Construct validity.}
	Gold localization is defined using reference modified-file sets. This is a practical oracle-style proxy for relevant edit scope, though not a perfect semantic oracle: some historical edits may be incidental, and some causally relevant files may remain unmodified. We therefore use gold localization as a diagnostic oracle-style intervention, not as an exact semantic file set.
	
	\textbf{Internal validity.}
	LLM-based agent behavior is stochastic, the results are therefore single realized outcomes per arm rather than repeated-trial estimates. To reduce this threat, all comparisons are paired by instance ID under a shared runtime and evaluation harness; efficiency analyses additionally use only rows with numeric telemetry in both compared arms.
	
	\textbf{External validity.}
	The study is limited to SWE-bench Verified, three repair backbones, and two localizers. Effect sizes may differ for other issue distributions, repository scales, or richer agent frameworks. Still, cross-backbone consistency suggests the conclusion is not tied to one model pairing.

	\section{Data and Artifact Availability}
	The Loc2Repair framework is publicly available. The repository includes the framework code, protocol files, normalized run summaries, localization predictions, paired comparison tables, and scripts used to reproduce the reported metrics. Repository: \url{https://github.com/mohammad-nour-alawad/Loc2Repair}
	
	\section{Conclusion}
		
	This paper presented \emph{Loc2Repair}, a modular framework for stage-aware evaluation of repository-grounded LLM repair, and used it to study file-level localization across three repair backbones on 500 SWE-bench Verified tasks. Explicit localization consistently improved resolved rate over baseline repair, with significant pooled gains for both predicted localizers and larger gains for gold guidance. Pooled analysis also showed lower mean elapsed time, although token usage remained model-dependent.
	
	These findings support two conclusions. First, file-level localization is a first-class repair lever for repository-grounded agents. Second, stage-aware evaluation is necessary for understanding which component changes actually improve end-to-end repair. At the same time, the remaining unresolved volume under gold localization shows that better file targeting does not exhaust the problem. Future work should extend Loc2Repair to broader localizer--repair comparisons, richer staged repair pipelines, and repeated-run stability analysis across broader benchmarks.
	
	\subsection{Future Work}
	
	Future work should extend Loc2Repair along three directions:
	\begin{itemize}
		\item \textbf{Broader comparisons.} Evaluate more localizers, repair backbones, and localization-injection strategies, including harder and softer forms of file guidance.
		\item \textbf{Richer staged repair.} Decompose repair beyond localization and patch generation, with specialized steps for context selection, patch planning, failure diagnosis, and patch refinement.
	\end{itemize}
	
	\begin{acknowledgments}
	This work is supported by the Ministry of Economic Development of the Russian Federation (IGK 000000C313925P4C0002), agreement No139-15-2025-010.
	\end{acknowledgments}
	
	\section*{Declaration on Generative AI}
	The authors used ChatGPT for English wording, grammar, and typo correction. The authors reviewed the final text and take full responsibility for the paper's content.
	
	\bibliography{software}

@inproceedings{SahaLKP2013ASE_BLUiR,
	author = {Saha, Ripon K. and Lease, Matthew and Khurshid, Sarfraz and Perry, Dewayne E.},
	title = {Improving bug localization using structured information retrieval},
	year = {2013},
	isbn = {9781479902156},
	publisher = {IEEE Press},
	url = {https://doi.org/10.1109/ASE.2013.6693093},
	doi = {10.1109/ASE.2013.6693093},
	booktitle = {Proceedings of the 28th IEEE/ACM International Conference on Automated Software Engineering},
	pages = {345–355},
	numpages = {11},
	location = {Silicon Valley, CA, USA},
	series = {ASE '13}
}

@inproceedings{YeBL2014FSE_LTR,
	author = {Ye, Xin and Bunescu, Razvan and Liu, Chang},
	title = {Learning to rank relevant files for bug reports using domain knowledge},
	year = {2014},
	isbn = {9781450330565},
	publisher = {Association for Computing Machinery},
	address = {New York, NY, USA},
	url = {https://doi.org/10.1145/2635868.2635874},
	doi = {10.1145/2635868.2635874},
	booktitle = {Proceedings of the 22nd ACM SIGSOFT International Symposium on Foundations of Software Engineering},
	pages = {689–699},
	numpages = {11},
	location = {Hong Kong, China},
	series = {FSE 2014}
}

@article{WangLo2016AmalgamPlus,
	author = {Wang, Shaowei and Lo, David},
	title = {AmaLgam+: Composing Rich Information Sources for Accurate Bug Localization},
	journal = {Journal of Software: Evolution and Process},
	volume = {28},
	number = {10},
	pages = {921-942},
	doi = {10.1002/smr.1801},
	url = {https://onlinelibrary.wiley.com/doi/abs/10.1002/smr.1801},
	eprint = {https://onlinelibrary.wiley.com/doi/pdf/10.1002/smr.1801},
	year = {2016}
}

@inproceedings{AkbarK2020MSR_Ablation,
	author = {Akbar, Shayan A. and Kak, Avinash C.},
	title = {A Large-Scale Comparative Evaluation of IR-Based Tools for Bug Localization},
	year = {2020},
	isbn = {9781450375177},
	publisher = {Association for Computing Machinery},
	address = {New York, NY, USA},
	url = {https://doi.org/10.1145/3379597.3387474},
	doi = {10.1145/3379597.3387474},
	booktitle = {Proceedings of the 17th International Conference on Mining Software Repositories},
	pages = {21–31},
	numpages = {11},
	location = {Seoul, Republic of Korea},
	series = {MSR '20}
}

@article{Niu2025CSUR_DLmeetsIR,
	author = {Niu, Feifei and Li, Chuanyi and Liu, Kui and Xia, Xin and Lo, David},
	title = {When Deep Learning Meets Information Retrieval-based Bug Localization: A Survey},
	year = {2025},
	issue_date = {November 2025},
	publisher = {Association for Computing Machinery},
	address = {New York, NY, USA},
	volume = {57},
	number = {11},
	issn = {0360-0300},
	url = {https://doi.org/10.1145/3734217},
	doi = {10.1145/3734217},
	journal = {ACM Comput. Surv.},
	month = jun,
	articleno = {296},
	numpages = {41},
}

@inproceedings{SWEbench2024,
	title={{SWE}-bench: Can Language Models Resolve Real-world Github Issues?},
	author={Carlos E Jimenez and John Yang and Alexander Wettig and Shunyu Yao and Kexin Pei and Ofir Press and Karthik R Narasimhan},
	booktitle={The Twelfth International Conference on Learning Representations},
	year={2024},
	url={https://openreview.net/forum?id=VTF8yNQM66}
}

@inproceedings{SWEagent2024,
	title={{SWE}-agent: Agent-Computer Interfaces Enable Automated Software Engineering},
	author={John Yang and Carlos E Jimenez and Alexander Wettig and Kilian Lieret and Shunyu Yao and Karthik R Narasimhan and Ofir Press},
	booktitle={The Thirty-eighth Annual Conference on Neural Information Processing Systems},
	year={2024},
	url={https://openreview.net/forum?id=mXpq6ut8J3}
}

@article{Agentless2025,
	author = {Xia, Chunqiu Steven and Deng, Yinlin and Dunn, Soren and Zhang, Lingming},
	title = {Demystifying LLM-Based Software Engineering Agents},
	year = {2025},
	issue_date = {July 2025},
	publisher = {Association for Computing Machinery},
	address = {New York, NY, USA},
	volume = {2},
	number = {FSE},
	url = {https://doi.org/10.1145/3715754},
	doi = {10.1145/3715754},
	journal = {Proc. ACM Softw. Eng.},
	month = jun,
	articleno = {FSE037},
	numpages = {24},
}

@inproceedings{LocAgent2025,
	author    = {Zhaoling Chen and Robert Tang and Gangda Deng and Fang Wu and Jialong Wu and Zhiwei Jiang and Viktor Prasanna and Arman Cohan and Xingyao Wang},
	title     = {LocAgent: Graph-Guided {LLM} Agents for Code Localization},
	booktitle = {Proceedings of the 63rd Annual Meeting of the Association for Computational Linguistics (Volume 1: Long Papers)},
	address   = {Vienna, Austria},
	publisher = {Association for Computational Linguistics},
	pages     = {8697--8727},
	year      = {2025},
	isbn      = {979-8-89176-251-0},
	doi       = {10.18653/v1/2025.acl-long.426},
	url       = {https://aclanthology.org/2025.acl-long.426/}
}

@inproceedings{AlAwadPrefiltering2025,
	author    = {Al Awad, Mohammad Nour and Ivanov, Sergey and Tikhonova, Olga},
	title     = {Pre-Filtering Code Suggestions using Developer Behavioral Telemetry to Optimize LLM-Assisted Programming},
	booktitle = {Proceedings of the 40th IEEE/ACM International Conference on Automated Software Engineering Workshops (ASEW)},
	year      = {2025},
	pages     = {113--120},
	doi       = {10.1109/ASEW67777.2025.00032}
}

@article{CoSIL2025,
	author        = {Zhonghao Jiang and Xiaoxue Ren and Meng Yan and Wei Jiang and Yong Li and Zhongxin Liu},
	title         = {CoSIL: Software Issue Localization via {LLM}-Driven Code Repository Graph Searching},
	journal       = {CoRR},
	volume        = {abs/2503.22424},
	year          = {2025},
	eprint        = {2503.22424},
	archivePrefix = {arXiv},
	primaryClass  = {cs.SE},
	url           = {https://arxiv.org/abs/2503.22424}
}

@InProceedings{OrcaLoca2025,
	title = 	 {{O}rca{L}oca: An {LLM} Agent Framework for Software Issue Localization},
	author =       {Yu, Zhongming and Zhang, Hejia and Zhao, Yujie and Huang, Hanxian and Yao, Matrix and Ding, Ke and Zhao, Jishen},
	booktitle = 	 {Proceedings of the 42nd International Conference on Machine Learning},
	pages = 	 {73416--73436},
	year = 	 {2025},
	volume = 	 {267},
	series = 	 {Proceedings of Machine Learning Research},
	month = 	 {13--19 Jul},
	publisher =    {PMLR},
	url = 	 {https://proceedings.mlr.press/v267/yu25x.html},
}

@misc{GottaCatchEmAll2025,
	title={Gotta catch 'em all! Towards File Localisation from Issues at Large}, 
	author={Jesse Maarleveld and Jiapan Guo and Daniel Feitosa},
	year={2025},
	eprint={2507.18319},
	archivePrefix={arXiv},
	primaryClass={cs.SE},
	url={https://arxiv.org/abs/2507.18319}, 
}

@misc{RGFL2026,
	title={RGFL: Reasoning Guided Fault Localization for Automated Program Repair Using Large Language Models}, 
	author={Melika Sepidband and Hamed Taherkhani and Hung Viet Pham and Hadi Hemmati},
	year={2026},
	eprint={2601.18044},
	archivePrefix={arXiv},
	primaryClass={cs.SE},
	url={https://arxiv.org/abs/2601.18044}, 
}

@misc{SGAgent2026,
	title={SGAgent: Suggestion-Guided LLM-Based Multi-Agent Framework for Repository-Level Software Repair}, 
	author={Quanjun Zhang and Chengyu Gao and Yu Han and Ye Shang and Chunrong Fang and Zhenyu Chen and Liang Xiao},
	year={2026},
	eprint={2602.23647},
	archivePrefix={arXiv},
	primaryClass={cs.SE},
	url={https://arxiv.org/abs/2602.23647}, 
}

@inproceedings{SweRank2026,
	title={{SWER}ank: Software Issue Localization with Code Ranking},
	author={Revanth Gangi Reddy and Tarun Suresh and JaeHyeok Doo and Ye Liu and Xuan-Phi Nguyen and Yingbo Zhou and Semih Yavuz and Caiming Xiong and Heng Ji and Shafiq Joty},
	booktitle={The Fourteenth International Conference on Learning Representations},
	year={2026},
	url={https://openreview.net/forum?id=OnkRqbNhe3}
}

@article{DEVLoRe2025,
	author = {Feng, Qiong and Ma, Xiaotian and Sheng, Jiayi and Feng, Ziyuan and Song, Wei and Liang, Peng},
	title = {Integrating Various Software Artifacts for Better LLM-based Bug Localization and Program Repair},
	year = {2025},
	publisher = {Association for Computing Machinery},
	address = {New York, NY, USA},
	issn = {1049-331X},
	url = {https://doi.org/10.1145/3770581},
	doi = {10.1145/3770581},
	note = {Just Accepted},
	journal = {ACM Trans. Softw. Eng. Methodol.},
	month = oct,
}

@misc{MiniSWEAgent2025,
	author       = {{SWE-agent Team}},
	title        = {{mini-SWE-agent}},
	howpublished = {\url{https://github.com/SWE-agent/mini-swe-agent}},
	year 		 = {2024},
	note         = {Accessed: 2026-05-03}
}

@inproceedings{AlAwadTiming2025,
	author    = {Al Awad, Mohammad Nour and Ivanov, Sergey and Tikhonova, Olga},
	title     = {Optimizing LLM Code Suggestions: Feedback-Driven Timing with Lightweight State Bounds},
	booktitle = {Proceedings of the 40th IEEE/ACM International Conference on Automated Software Engineering Workshops (ASEW)},
	year      = {2025},
	pages     = {213--220},
	doi       = {10.1109/ASEW67777.2025.00049}
}

@inproceedings{AutoCodeRover2024,
	author = {Zhang, Yuntong and Ruan, Haifeng and Fan, Zhiyu and Roychoudhury, Abhik},
	title = {AutoCodeRover: Autonomous Program Improvement},
	year = {2024},
	isbn = {9798400706127},
	url = {https://doi.org/10.1145/3650212.3680384},
	doi = {10.1145/3650212.3680384},
	booktitle = {Proceedings of the 33rd ACM SIGSOFT International Symposium on Software Testing and Analysis},
	pages = {1592–1604},
	numpages = {13},
	series = {ISSTA 2024}
}

@inproceedings{RepairAgent2024,
	author = {Bouzenia, Islem and Devanbu, Premkumar and Pradel, Michael},
	title = {RepairAgent: An Autonomous, LLM-Based Agent for Program Repair},
	year = {2025},
	isbn = {9798331505691},
	publisher = {IEEE Press},
	url = {https://doi.org/10.1109/ICSE55347.2025.00157},
	doi = {10.1109/ICSE55347.2025.00157},
	booktitle = {Proceedings of the IEEE/ACM 47th International Conference on Software Engineering},
	pages = {2188–2200},
	numpages = {13},
	series = {ICSE '25}
}

@inproceedings{OpenHands2025,
	author    = {Wang, Xingyao and Li, Boxuan and Song, Yufan},
	title     = {{OpenHands}: An Open Platform for {AI} Software Developers as Generalist Agents},
	booktitle = {The Thirteenth International Conference on Learning Representations},
	year      = {2025},
	url       = {https://openreview.net/forum?id=OJd3ayDDoF}
}

@inproceedings{ThinkRepair2024,
	author = {Yin, Xin and Ni, Chao and Wang, Shaohua and Li, Zhenhao and Zeng, Limin and Yang, Xiaohu},
	title = {ThinkRepair: Self-Directed Automated Program Repair},
	year = {2024},
	isbn = {9798400706127},
	url = {https://doi.org/10.1145/3650212.3680359},
	doi = {10.1145/3650212.3680359},
	booktitle = {Proceedings of the 33rd ACM SIGSOFT International Symposium on Software Testing and Analysis},
	pages = {1274–1286},
	numpages = {13},
	series = {ISSTA 2024}
}

@misc{FixAgent2024,
      title={UniDebugger: Hierarchical Multi-Agent Framework for Unified Software Debugging}, 
      author={Cheryl Lee and Chunqiu Steven Xia and Longji Yang and Jen-tse Huang and Zhouruixin Zhu and Lingming Zhang and Michael R. Lyu},
      year={2025},
      eprint={2404.17153},
      archivePrefix={arXiv},
      primaryClass={cs.SE},
      url={https://arxiv.org/abs/2404.17153}, 
}

@inproceedings{RepoDebug2025,
    title = "{R}epo{D}ebug: Repository-Level Multi-Task and Multi-Language Debugging Evaluation of Large Language Models",
    author = "Liu, Jingjing  and
      Liu, Zeming  and
      Cheng, Zihao  and
      He, Mengliang  and
      Shi, Xiaoming  and
      Guo, Yuhang  and
      Zhu, Xiangrong  and
      Guo, Yuanfang  and
      Wang, Yunhong  and
      Wang, Haifeng",
    booktitle = "Findings of the Association for Computational Linguistics: EMNLP 2025",
    month = nov,
    year = "2025",
    publisher = "Association for Computational Linguistics",
    url = "https://aclanthology.org/2025.findings-emnlp.1294/",
    doi = "10.18653/v1/2025.findings-emnlp.1294",
    pages = "23784--23813",
    ISBN = "979-8-89176-335-7",
}

@misc{SWEPolyBench2025,
	title={{SWE}-PolyBench: A multi-language benchmark for repository level evaluation of coding agents},
	author={Muhammad Shihab Rashid and Christian Bock and Yuan Zhuang and Alexander Buchholz and Timothy B Esler and Simon Valentin and Luca Franceschi and Martin Wistuba and Prabhu Teja S and Woojung Kim and Anoop Deoras and Giovanni Zappella and Laurent Callot},
	year={2026},
	url={https://openreview.net/forum?id=n577FC6CKk}
}

@misc{SWESynth2025,
	author        = {Pham, Minh V. T. and Phan, Huy N. and Phan, Hoang N. and Chi, Cuong Le and Nguyen, Tien N. and Bui, Nghi D. Q.},
	title         = {{SWE-Synth}: Synthesizing Verifiable Bug-Fix Data to Enable Large Language Models in Resolving Real-World Bugs},
	year          = {2025},
	eprint        = {2504.14757},
	archivePrefix = {arXiv},
	primaryClass  = {cs.SE},
	doi           = {10.48550/arXiv.2504.14757},
	url           = {https://arxiv.org/abs/2504.14757}
}

@misc{ExpeRepair2025,
	author        = {Mu, Fangwen and Wang, Junjie and Shi, Lin and Wang, Song and Li, Shoubin and Wang, Qing},
	title         = {{ExpeRepair}: Dual-Memory Enhanced {LLM}-based Repository-Level Program Repair},
	year          = {2025},
	eprint        = {2506.10484},
	archivePrefix = {arXiv},
	primaryClass  = {cs.SE},
	doi           = {10.48550/arXiv.2506.10484},
	url           = {https://arxiv.org/abs/2506.10484}
}

@article{Hossain2024DeepDive,
	author = {Hossain, Soneya Binta and Jiang, Nan and Zhou, Qiang and Li, Xiaopeng and Chiang, Wen-Hao and Lyu, Yingjun and Nguyen, Hoan and Tripp, Omer},
	title = {A Deep Dive into Large Language Models for Automated Bug Localization and Repair},
	year = {2024},
	issue_date = {July 2024},
	volume = {1},
	number = {FSE},
	url = {https://doi.org/10.1145/3660773},
	doi = {10.1145/3660773},
	journal = {Proc. ACM Softw. Eng.},
	month = jul,
	articleno = {66},
	numpages = {23},
}

\end{document}